\documentclass[prl,superscriptaddress,twocolumn,showpacs,preprintnumbers,%
amsmath,amssymb]{revtex4}

\usepackage{graphicx}
\usepackage{dcolumn}
\usepackage{bm}
\usepackage{amsbsy}

\begin{document}   
\title{Theory of the hourglass dispersion of magnetic 
excitations in high-T$_c$ cuprates} 
\author{Roland Zeyher}
\affiliation{Max-Planck-Institut f\"ur Festk\"orperforschung,
             Heisenbergstrasse 1, D-70569 Stuttgart, Germany}

\date{\today}

\begin{abstract}
A theory for the dispersion of collective magnetic excitations in
superconducting cuprates is presented with the aim to cover both 
high and low doping regimes. Besides of spin fluctuations
describable in the random phase approximation (RPA) we allow for
local spin rotations within a mode-coupling theory. At low temperatures 
and moderately large correlation lengths we obtain two branches of 
excitations which disperse up- and downwards 
exhibiting the hourglass behavior observed experimentally at intermediate 
dopings. At large and small dopings our theory essentially reduces to
the RPA and spin wave theory, respectively.
\end{abstract}


\maketitle


The low-temperature magnetic response of many high-T$_c$ superconductors is 
characterized by a resonant mode inside the superconducting gap
around the antiferromagnetic wave vector $\bf Q$. 
This collective mode manifests itself as a single peak at $\bf Q$ 
which splits into two peaks dispersing up- and downwards in frequency
away from $\bf Q$. This unusual dispersion resembles the shape of a 
hourglass\cite{Fau,Hin}. 
Theories to explain this phenomena use either a
more local\cite{Sega} or an itinerant\cite{Norman,Yamase} description
of the magnetism.    
The second approach considers particle-hole excitations
with spin flips which interact within the random phase
approximation (RPA) forming a dispersing bound state in the superconducting 
gap.
This approach yields only one branch of excitations below the Stoner
continuum whereas it has been established recently that the lower branch, the
center of the hourglass as well as part of the upper branch lie below this 
continuum in the gapped region\cite{Fau}.
A more theoretical argument for the incompleteness of a RPA description 
comes from the fact that different spin directions do not mix as a function
of time in this approximation which excludes local rotations of spins 
known from spin wave theory.  
Below we will present a theory which contains both spin wave 
theory and RPA as special cases. At intermediate 
dopings we will show that both RPA and spin wave like spin fluctuations 
are important and produce the two branches of the hourglass dispersion.  

We consider the $t$-$J$ model\cite{Ogata} with the Hamiltonian $H$,
\begin{equation}
H =
\sum_{{\bf k}\sigma}\epsilon({\bf k}) {\tilde c}^\dagger_{{\bf k}\sigma}
{\tilde c}_{{\bf k}\sigma} + \frac{1}{2} \sum_{\bf k} J({\bf k})
{\bf S}_{\bf k} {\bf S}_{-\bf{k}}.
\label{H}
\end{equation}
${\tilde c}^\dagger_{{\bf k}\sigma}, {\tilde c}_{{\bf k}\sigma}$ are creation and 
annihilation operators, respectively, for electrons with momentum $\bf k$ and
spin projection $\sigma$ excluding any double occupancies of sites. 
${\bf S}_{\bf k}$ are
spin operators in momentum space, $\epsilon({\bf k})$ and
$J({\bf k})$ are the bare electron dispersion and the Fourier transform of the 
Heisenberg coupling, respectively. A useful approximation 
for $H$, used in the following, is obtained by 
taking the large $N$ limit of the $t$-$J$ model, where $\epsilon({\bf k})$
describes a renormalized dispersion of quasi-particles and the fermionic 
operators can be treated as usual creation and annihilation operators.  

In the following we are interested in the time evolution of the
spin operator ${\bf S}_{\bf k} = 1/2 \sum_{\bf q} {\bf A}({\bf k},{\bf q})$,
\begin{equation}
{\bf A}({\bf k},{\bf q}) = 
\sum_{\alpha,\beta} c^\dagger_{{{\bf k}+{\bf q}}\alpha}
{\mbox{\boldmath $\sigma$ \unboldmath}\hspace{-0.15cm}}_{\alpha\beta} 
c_{{\bf q}\beta},
\label{A}
\end{equation}
where 
\boldmath $\sigma$ \unboldmath
denotes the vector of the three Pauli matrices.
It obeys the equation of motion 
\begin{equation}
\frac{\partial {\bf A}}{\partial t} 
=i({\cal L}_0+{\cal L}_1){\bf A},
\end{equation}
with
\begin{eqnarray}
{\cal L}_0 {\bf A}({\bf k},{\bf q}) = (\epsilon({\bf k}+{\bf q})
-\epsilon({\bf q})){\bf A}({\bf k},{\bf q}) \nonumber \\
+J({\bf k}) (f({\bf q})-f({\bf k}+{\bf q}))
{\bf S}_{\bf k},
\label{L1}
\end{eqnarray}

\begin{eqnarray}
{\cal L}_1 {\bf A}({\bf k},{\bf q}) = -\frac{i}{2} 
\sum_{{\bf k}'} J({{\bf k}'})\cdot  \nonumber \\    
\Bigl({\bf S}_{{\bf k}'} \times  ({\bf A}({\bf k}-{\bf k'},{\bf q})
+{\bf A}({\bf k}-{\bf k'},{\bf k'}+{\bf q})) \Bigr).
\label{L2}
\end{eqnarray}
$f({\bf k})$ is equal to $\langle c^\dag_{{\bf k}\sigma} 
c_{{\bf k}\sigma} \rangle$,
where $\langle ...\rangle$ denotes the thermodynamic expectation value, 
and $\times$ stands for the vector product. Since we are only interested
in the spin response we have dropped terms on the right-hand side of 
Eq.(\ref{L2}) which involve fluctuations in the density. We also
dropped an overall prefactor denoting the number of primitive cells.  
The unperturbed Liouville operator ${\cal L}_0$ describes the time
evolution of the system in the RPA. From its explicit expression in
Eq.(\ref{L1}) follows that it does not mix different cartesian 
components of the spin operators.
In contrast to that the time evolution described by ${\cal L}_1$ involves
product states of spin operators, mixes different spin 
components and thus can describe rotations of spins due to fluctuating
fields.  

The spin susceptibility $\chi({\bf k},z)$
can conveniently be calculated from the associated Kubo
relaxation function $\Phi({\bf k},z) = (\chi({\bf k}) -
\chi({\bf k},z))/z$ where $z$ is a complex frequency and 
$\chi({\bf k})$ is equal to $\chi({\bf k},z=0)$. Due to the rotational
invariance in spin space we may assume that $\chi$, $M$ etc. always
refer to the z-direction.
Using the Mori formalism $\Phi$ can be written as\cite{Forster}
\begin{equation}
\Phi({\bf k},z) = \frac{\chi({\bf k})}{z+M^{\rm RPA}({\bf k},z)
+M({\bf k},z)}.
\label{Phi1}
\end{equation} 
The first memory kernel $M^{\rm RPA}({\bf k},z)$ describes the time evolution of
spin operators by ${\cal L}_0$. According to Eq.(\ref{L1}) 
the direction of the spin operators is conserved and they remain
always linear in the operators ${\bf A}$. Eliminating the ${\bf A}$
operators in the equation of motion in favor of the orginal ${\bf S}$
operators yields an explicit expression for $M^{\rm RPA}$ which may be expressed
in terms of the RPA spin susceptibility  $\chi^{\rm RPA}({\bf k},z)$,   
\begin{equation}
M^{\rm RPA}({\bf k},z) = z \chi^{\rm RPA}({\bf k},z)/(\chi^{\rm RPA}({\bf k})
-\chi^{\rm RPA}({\bf k},z)),
\label{RPA}
\end{equation}
with
\begin{equation}
1/\chi^{\rm RPA}({\bf k},\omega) = 1/\chi^{(0)}({\bf k},\omega)
+J({\bf k}),
\label{RPA0}
\end{equation}
$\chi^{(0)}({\bf k},\omega)$ is the free susceptibility.

The second memory contribution $M({\bf k},z)$ is due to the time evolution
of single spin operators ${\bf S}$ into product states of spin operators
described by ${\cal L}_1$. Using the mode-coupling assumption 
and performing the analytic
continuation $z \rightarrow \omega +i \eta$, we obtain for the
imaginary part of $M$, denotd by $M''$,
\begin{equation}
M''({\bf k},\omega) =  \sum_{\bf k'} (J({\bf k}')-J({\bf k}-{\bf k}'))^2
D''({\bf k},{\bf k'},\omega)/(\omega \chi({\bf k})),
\label{M''}
\end{equation}   
\begin{eqnarray}
D''({\bf k},{\bf k'},\omega) = \pi \int d \omega' 
A({\bf k}-{\bf k'},\omega -\omega') A({\bf k'},\omega') \nonumber \\ 
(b({\omega'})-b({\omega'}-\omega)).
\label{D}
\end{eqnarray}
$A$ is the spectral function of the spin
propagator and $b(\omega)$ the Bose function. 
\section{Small correlation lengths}
In the case of
small antiferromagnetic correlation lengths $\xi$, corresponding
to the overdoped regime, the RPA should be a reasonable approximation 
for the spin susceptibility.
The dashed line in Fig. 1 shows the imaginary part of 
$\chi^{\rm RPA}({\bf Q},\omega)$
for ${\bf Q}=({\pi,\pi})$  using the
parameters tb2 in Table I of Ref.\cite{Norman} and 
a chemical potential corresponding to the doping $\delta=0.20$.
The energy unit is 1 eV in the following and lengths are measured in units
of the lattice constant a of the square lattice. The superconducting
order parameter is $\Delta({\bf k}) = \Delta (\cos k_x -\cos k_y)/2$ with
$\Delta = 0.029$, $J$ equal to 0.135 and $\eta=0.004$.
The dashed line in Fig. 1 illustrates that most of
the spectral weight resides in the bound state at the energy 0.038 and that 
only a small part
of it has been left in the continuum at higher energies.     
Away from $(\pi,\pi)$ the dashed curve in Fig. 1 does not 
change dramatically as long as
the bound state lies still in the gapped region. Entering the particle-hole
continuum by going further away from $(\pi,\pi)$ destroys the bound state
and most of the spectral weight shifts to high energies of the order of $t$.
\begin{figure}
\vspace{-0.3cm}
\includegraphics[angle=270,width=8cm]{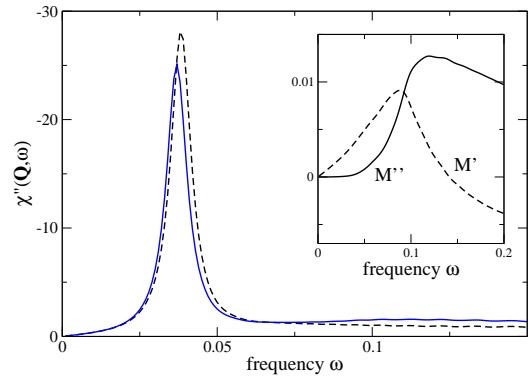}
\caption{\label{fig:1}
$\chi''({\bf Q},\omega)$ without (dashed line)
and with (solid line) memory function $M$ for a
doping $\delta = 0.20$ and $T=0$. The dashed curve corresponds to the RPA.
Inset: real part $M'$ and imaginary part $M''$ of 
$M({\bf Q},\omega)$.
}
\vspace{-0.5cm}
\end{figure}
The parameters used in Fig. 1 yield $\xi \sim 0.78$.
For such a small $\xi$ practically all momenta in the 
sum over momenta in Eq.(\ref{M''}) contribute substantially which 
means that $M''$ is mainly determined by contributions away from the 
small region around $(\pi,\pi)$ so that the bound state 
and its low-energy scale is rather irrelevant for $M$.
This is
confirmed by an explicit calculation of $M$ using RPA results for
the various quantities in Eq.(\ref{M''}). The result is
shown in the inset of Fig. 1 for T=0. 
$M''$ (solid line) is structureless except at small energies where it vanishes
rapidly due to the smallness of $A$ in this region and the cutoff $\omega$
for the integration over $\omega'$ in Eq.(\ref{D}) due to the bosonic factors.
Taking $M$ into account in Eq.(\ref{Phi1})
yields the solid line in Fig. 1 which differs only 
marginally from the dashed line. This shows
that at short correlation lengths the RPA result for $\chi''$
is essentially correct and that the correction $M$ to $M^{\rm RPA}$
is rather small. The underlying physical picture is that the momentary local
axis of preferred spin directions fluctuates very rapidly due to the
random forces induced by ${\cal L}_1$. The spectrum of these forces is
given by $M"$ and characterized by the large energy scale $t$ in agreement with
the inset of Fig. 1.
\section{Large correlation lengths}
For large $\xi$ the spectral function $A({\bf k'},\omega')$
is strongly peaked at ${\bf k'}={\bf Q}$. This means that  
the integration 
over $\bf k'$ in Eq.(\ref{M''}) is restricted to momenta 
near $\bf 0$ or near $\bf Q$.  Since we are interested in external momenta
${\bf k} \sim {\bf Q}$ the momentum of one of the two spectral functions 
in Eq.(\ref{D}) is small. Due to spin conservation this spectral
function describes spin diffusion and is mainly restricted to small
values of $\omega'$. As a result one may neglect the small frequency transfer
in the second spectral function in Eq.(\ref{D}). Taking also the real part
of $M$ into account we obtain from Eqs.(\ref{M''}) and (\ref{D}), 
\begin{equation}
M({\bf k},\omega) = -\omega^2({\bf k}) \Phi({\bf Q},\omega)/\chi({\bf Q}),
\label{M}
\end{equation}
with 
\begin{equation}
\omega^2({\bf k}) = \frac{2}{\chi(\bf k)} \sum_{\bf q} 
(J({\bf q})-J({\bf k}-{\bf q}))^2
\langle S_{{\bf k}-{\bf q}} S_{{\bf q}-{\bf k}} \rangle \chi({\bf q})),
\label{omega}
\end{equation}
and the equal-time correlation function 
\begin{equation}
\langle S_{\bf k}  S_{\bf -k} \rangle = 
\int d\omega b({\omega}) A({\bf k},\omega). 
\label{equal}
\end{equation}
In deriving Eq.(\ref{M}) we used the fact that the two memory functions
in Eq.(\ref{Phi1}) depend for our parameters only slowly on momentum 
around the wave vector ${\bf Q}$ so that the combination $\Phi/\chi$ on
the right-hand side of Eq.(\ref{M}) may be evaluated at $\bf Q$.  
The sum over $\bf q$ in Eq.(\ref{omega}) runs over half of the Brillouin
zone centered around $\bf Q$. 
The evaluation of the above expressions
using the RPA encounters a problem: $\xi$, calculated in the
RPA, is in the  optimal and moderately underdoped region around one 
or smaller and
increases substantially only near the transition to the antiferromagnetic
state in disagreement with the experiment. For instance, we have 
for $\delta=0.12$ $\xi \sim 0.8$, Ref.\cite{Yamase} $\xi \sim 0.6$
using quite different parameter values,
whereas the experimental values for $\xi$ are larger by about a factor 5 or 
more\cite{Dai}.
Since this large discrepancy would affect severely the momentum sum in
Eq.(\ref{omega}) we prefer to use a realistic 
$\chi({\bf k})$ as input in calculating $M$ and write
$\chi({\bf k}) = \chi({\bf Q})/(1+\xi^2(\Delta {\bf k})^2)$ for
$\Delta {\bf k} \equiv {\bf k}-{\bf Q} \sim 0 $ considering $\xi$ as a 
parameter to be determined from experiment.
\vspace{-0.3cm}
\begin{figure}
\includegraphics[angle=270,width=8cm]{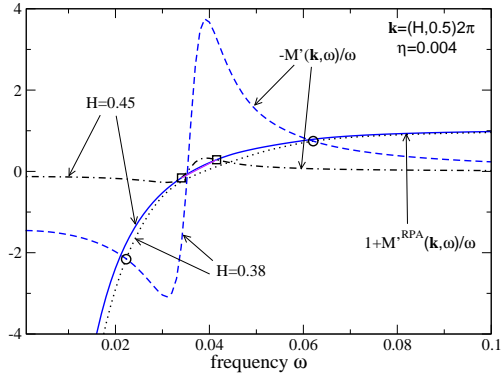}
\caption{\label{fig:2}
(color online)
$1+M'^{\rm RPA}({\bf Q},\omega)/\omega$ (solid and dotted lines)
and $-M'({\bf Q},\omega)/\omega$ (dash-dotted and dashed lines)
as a function of frequency for two momenta H. Squares and circles
denote poles of $\chi({\bf k},\omega)$. 
}
\vspace{-0.5cm}
\end{figure}
\begin{figure}
\includegraphics[angle=270,width=6cm]{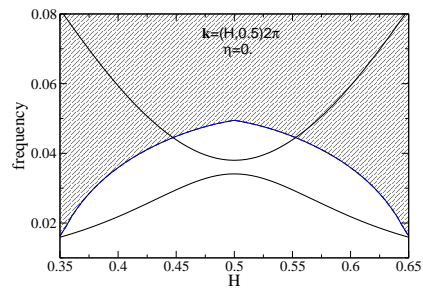}
\caption{\label{fig:3}
(color online)
Position of the poles of $\chi({\bf k},\omega)$ 
as a function of momentum. The shadowed region marks the 
particle-hole continuum.
}
\vspace{-0.7cm}
\end{figure}

It is instructive to study the frequency dependence of the denominator of 
Eq.(\ref{Phi1}). In order to describe a slightly underdoped case 
we choose the same parameters as in Fig. 1, a chemical potential
corresponding to $\delta = 0.12$, $\xi =5$, and the cutoff $1/\xi$
for the sum over ${\bf q}$ in Eq.(\ref{omega}). The solid and dotted line in 
Fig. 2 show $1 + M'^{\rm RPA}({\bf k},\omega)/\omega$ for H=0.5 and 0.38,
respectively, writing ${\bf k}=(H,0.5)2\pi$. This quantity is practically
independent of momentum, increases monotonically with $\omega$ and is
zero at the RPA resonance energy $\omega_R \sim 0.038$. The dashed and
dash-dotted lines in Fig. 2 show $-M'({\bf k},\omega)/\omega$ for the same 
momenta.
These curves resemble the real part of an oscillator located at $\omega_R$
with an oscillator strength being very small at $\Delta {\bf k} \equiv
{\bf k}-{\bf Q} =0$ and strongly increasing with $|\Delta {\bf k}|$.
The poles of Eq.(\ref{Phi1}) are given by the common points of the two 
curves denoted by squares and circles. Since the common point at 
$\omega = \omega_R$ (not shown in Fig. 2) has vanishing pole strength 
there are two branches of collective
spin excitations. For vanishing damping $\eta$ their dispersion is shown 
in Fig. 3 by solid lines.
They approximately touch each other at $\Delta {\bf k} =0$ and disperse 
up- and downwards with increasing $|\Delta {\bf k}|$. 
For not too large $|\Delta {\bf k}|$ both branches lie below the continuum 
in agreement with experiment\cite{Fau}. 
Performing the calculation in the normal state at $T=T_c$
the solid and dotted lines in Fig. 2 lie everywhere above zero but the
solid and dashed and also the dotted and dash-dotted lines have still one
common point at larger frequencies. In this case only the upper but
not the lower branch exists in agreement with 
experiment\cite{Hin}. At very low dopings $M^{\rm RPA} \rightarrow 0$
due to the constraint and the pole condition $\omega + M'({\bf k},\omega)$
yields in the presence of long-range order the correct spin wave 
dispersion\cite{Forster}.

Several prerequisites are necessary to obtain the above hourglass 
dispersion for spin excitations. There must exist two different kinds
of spin excitations to account for the two branches. The first one
are RPA spin fluctuations where all induced spin moments have the same
direction. They may be characterized by the fact that the internal 
fields induced by the Heisenberg interaction conserves frequency, momentum
and spin direction which is a direct consequence of the one-mode behavior
of ${\cal L}_0$ in Eq.(\ref{L1}). The second one are local rotations of
spins under the influence of ${\cal L}_1$ in Eq.(\ref{L2}). In this case
a spin in z direction acquires in its time evolution also a component
in x direction due to the presence of a spin fluctuation in y direction.   
The pure form of the two kind of spin excitations
are obtained for $M=0$ and $M^{\rm RPA}=0$, respectively, and are
realized approximately at large and small dopings. In the hourglass regime
$M$ and $M^{\rm RPA}$ are of similar magnitude.
 
Another prerequisite for hourglass behavior is that $\xi$ is substantially
larger than 1. Only then is the momentum integration in Eq.(\ref{M''})
restricted to the resonance and the spin diffusion modes yielding   
oscillator-like behavior of $M$ near $\omega_R$. The up- and downwards
dispersion and their approximate degeneracy at $\bf Q$ is mainly 
determined by $\omega^2({\bf k})$, which according to Eq.(\ref{omega})
is roughly proportional to $\langle S_{\Delta {\bf k}}S_{-\Delta {\bf k}} \rangle
(\xi^{-2}+(\Delta {\bf k})^2)$. The first factor tends to zero at low
temperatures for $\Delta {\bf k} \rightarrow 0$ and saturates at large
$\Delta {\bf k}$. As a result $\omega^2({\bf k})$ is very small at 
$\Delta {\bf k} =0$ causing the approximate touching of the two branches
at $\omega=\omega_R$ and $\Delta {\bf k} = 0$. With increasing 
$\Delta {\bf k}$ $\omega^2({\bf k})$ increases strongly leading
to a downward dispersion of the lower branch even if $\omega_R$ was
practically dispersionless as in our case. According to Fig. 3 the upper
branch increases at large $|\Delta {\bf k}|$ roughly as 
$0.6 J |\Delta {\bf k}|$, i.e., with an effective spin wave velocity which 
is reduced by about a factor 2-3 compared to spin wave theory similar as
in experiment\cite{Hin,Fau}. 
\begin{figure}[t] 
\vspace{-0.3cm}
\includegraphics[angle=270,width=8cm]{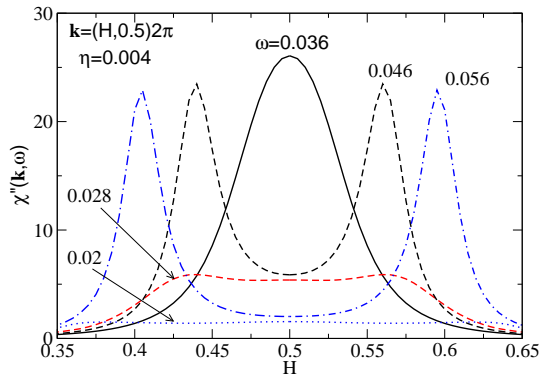}
\caption{\label{fig:4}
(color online)
Imaginary part of the spin susceptibility 
at zero temperature as a function of momentum for different frequencies. 
}
\vspace{-0.7cm}
\end{figure}
\begin{figure}[t]
\includegraphics[angle=270,width=8cm]{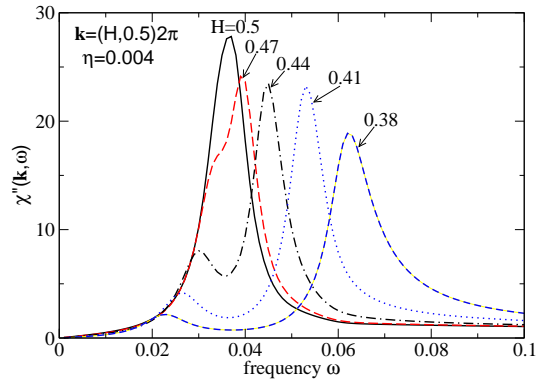}
\caption{\label{fig:5}
(color online)
Imaginary part of the spin susceptibility at zero temperature 
as a function of frequency for different momenta.
}
\vspace{-0.7cm}
\end{figure}

Using the same parameters as in Figs. 2 and 3 Fig. 4 shows $\chi''$
as a function of $H$ with the frequency as a parameter. As suggested by
Fig. 3 $\chi''$ exhibits a hourglass dispersion with intensities which
are largest near $\omega_R$ and decay rather fast and slow towards lower
and higher frequencies, respectively. Fig. 5 shows $\chi''$ as a function
of $\omega$ for a fixed $H$ as a parameter. In agreement with 
Fig. 4
the strong peak at $H=0.5$ splits into two peaks with decreasing $H$ which
disperse up- and downwards in frequency. The curve for $H=0.5$ 
calculated for the small damping $\eta = 0.004$ has the shape of a 
Lorentzian. At smaller dampings this peak splits into a double
peak due to the small gap between upper and lower branch shown in Fig. 3. 
For $M=0$ only the lower, weaker peak is obtained.

In conclusion, we have shown that the memory function
of the spin susceptibility contains in general two distinct contributions
due to RPA-like and due to rotational spin fluctuations. The first one
dominates at large, the second one at small dopings. At intermediate
dopings both are of similar magnitude which leads to one upwards and one 
downwards dispersing branch of excitations. At low temperatures the
two branches are approximately degenerate at $\bf Q$ which explains,
at least qualitatively, the 
observed hourglass dispersion at intermediate dopings.

The author is grateful to V. Hinkov, P. Horsch, D. Manske and H. Yamase 
for useful discussions. 
      

\end{document}